\documentclass[fleqn,10pt]{SelfArx} % Document font size and equations flushed left

\usepackage{lipsum} % Required to insert dummy text. To be removed otherwise

\usepackage{lscape}
\usepackage{tcolorbox}

\usepackage{soul}
\usepackage[square,sort,comma,numbers]{natbib}

\newcommand{\recommendation}[1]{
\begin{tcolorbox}[colback=gray!20,%gray background
                  colframe=color2,% black frame colour
                  width=8.5cm,
                  left=2mm,
                  right=2mm,
                  arc=2mm, auto outer arc,
                 ]
#1
\end{tcolorbox}}

\definecolor{color1}{RGB}{50,13,77} % Color of the article title and sections
\definecolor{color2}{RGB}{248,193,117} % Color of the boxes behind the abstract and headings

\usepackage{hyperref} % Required for hyperlinks

\hypersetup{
	hidelinks,
	colorlinks,
	breaklinks=true,
	urlcolor=color1,
	citecolor=color1,
	linkcolor=color1,
	bookmarksopen=false,
	pdftitle={Title},
	pdfauthor={Author},
}

\PaperTitle{Bringing together African \& European research communities with an inclusive astronomy conference} % Article title

\Authors{Chris M. Harrison\textsuperscript{1,$\star$}, Leah Morabito\textsuperscript{2,$\dagger$} and Ann Njeri\textsuperscript{1}, on behalf of the Organising Committees}

\affiliation{\textsuperscript{1}\textit{School of Mathematics, Statistics and Physics, Newcastle University, Newcastle upon Tyne, NE1 7RU, UK}} % Author affiliation
\affiliation{\textsuperscript{2}\textit{Centre for Extragalactic Astronomy, Department of Physics, Durham University, Durham DH1 3LE, UK}} % Author affiliation
\affiliation{$^{\star}$\textbf{Email}: christopher.harrison@newcastle.ac.uk, $^{\dagger}$\textbf{Email}: leah.k.morabito@durham.ac.uk} % Corresponding author

\Abstract{We report on an international scientific conference, where we brought together African and European academic astronomers. This aimed to bridge the gap between those in position of privilege, with ease of access to international events (i.e., the typical experience of academics in Western institutions), with those historically excluded (affecting the majority of African scientists/institutions). We describe how we designed the conference around cutting-edge research problems, but with a parallel focus on building networking and professional relationships. Significant effort went into: (1) ensuring a diversity of participants; (2) practically and financially supporting those who may never have attended an international conference and; (3) creating an inclusive and supportive environment through a careful programme of activities, both before and during the event. Maintaining scientific integrity was a core commitment throughout. We summarise successes, challenges and lessons learnt from organising this conference. We also present feedback obtained from participants immediately after the conference, and a discussion of some longer-term impacts, which we identified around 1 year later. We found an overall achievement of our objectives, and multiple longer-term benefits. With this report we provide some key recommendations for groups, from any research field, who may wish to lead similar initiatives.}

\begin{document}

\maketitle % Output the title and abstract box

%\tableofcontents % Output the contents section

\thispagestyle{empty} % Removes page numbering from the first page

%----------------------------------------------------------------------------------------
%	ARTICLE CONTENTS
%----------------------------------------------------------------------------------------

\section*{Motivation \& Scientific Context} 
\noindent Astronomy and space science can be used as an important tool for development and for achieving the
United Nations (UN) Sustainable Development Goals (SDGs)\footnote{\url{ https://sdgs.un.org/}} through: education; socio-economic growth via advances in science and technology; and promoting international peace and diplomacy \cite{McBride2018,Povic2021}. Partly driven by these development goals, astronomy and space science research is seeing significant growth on the continent of Africa. In recent years, a major contributing factor is the planning and development of the revolutionary radio observatory, the Square Kilometre Array (SKA)\footnote{\url{https://www.skao.in}}. This ambitious international project will be situated across Africa and Australia. Other major astronomy projects on the African continent include the South African Astronomical Observatory (SAAO) and the High Energy Stereoscopic System (H.E.S.S) observatory in Namibia\footnote{\url{https://www.mpi-hd.mpg.de/HESS/}} \citep[more examples in][]{Povic2021}. In parallel to these projects, the continent is seeing growth in postgraduate astronomy programmes, and international initiatives to support early-career African astronomers, such as the ``Development in Africa with Radio Astronomy'' project \citep[DARA,][]{Hoare2018} and other strategic partnerships (e.g., between the United Kingdom's Science and Technology Facilities Council [STFC], and the South African National Research Foundation\footnote{\url{https://www.ukri.org/news/forging-a-new-partnership-with-south-africa/}}).

Despite this positive growth of projects and initiatives, developing astronomy research in Africa faces many on-going challenges. These are shared across many research topics and institutions across the continent, and for people of African origin (but may be working in other countries). These relate to: limited opportunities and resources; the limited retention of young people into higher education; issues with equity and inclusion (including challenges around travel visas); and access to the latest international knowledge, facilities and networks \citep{Woolston2019,Makoni2023}. 

Motivated by addressing some of these these challenges, in July 2024, Durham University and Newcastle University jointly hosted an international astronomy conference: ``AGN Populations Across Continents and Cosmic Time''. The main goal of this conference was to hold a scientific meeting, discussing timely scientific problems, but with a broader impact to integrate African researchers into the international community.
Additionally, European researchers can often operate in their own continental spheres, and this conference offered the opportunity for them to broaden their networks. Therefore, this workshop aimed to facilitate networking that will strengthen ties between the European and African scientific communities.

\subsection*{Scientific context}
Supermassive black holes, with masses of millions to billions times that of the Sun, are located at the nuclei of galaxies. These black holes grow by the accretion of gas. As this gas falls towards the black holes, it becomes extremely bright. Furthermore, jets of charged particles can be launched in the vicinity of these accreting black holes, and these jets can extend over vast distances within the galaxies and beyond. When these accretion events are detected with astronomical observatories on Earth or in space, they are known as ``Active Galactic Nuclei'' (AGN). 

Both observational studies, and simulations, have shown that the energy released by AGN is important for galaxy evolution. However, there continue to be major research challenges in the field, including: (1) obtaining a complete census of AGN events; (2) understanding the detailed physical structure of the material associated with AGN; (3) establishing the properties of AGN host galaxies and the details of {\em how} AGN regulate galaxy growth, and; (4) identifying the key techniques and datasets required to make significant progress in answering these scientific questions over the coming decade \citep[review in e.g.,][]{Harrison2024}. The scientific scope of our conference was to cover these four main themes.\footnote{We note that these themes cover a number of the STFC's \href{https://www.ukri.org/publications/stfc-science-challenges/stfc-science-challenges-in-frontier-physics/}{Science Challenges in frontier physics (A1 - A5)}.} Due to the focus on the African scientific community, the workshop looked to particularly showcase work on this scientific topic that uses SKA pathfinder observatories in addition to other astronomical facilities based on the African continent.

%------------------------------------------------
\section*{Selecting and Supporting the Participants}

Careful planning went into selecting participants. During selection, our primary objectives were: (1) ensuring a scientifically productive meeting, with participants working on the relevant research problems, balanced across the four key themes (outlined above); and (2) having significant participation from those of African institutions and/or with African origin (aiming for $\approx$30--50\%). We detail the processes of selecting the organisational committees, choosing invited speakers and the selection of participants. 

\begin{figure*}[ht]\centering % Using \begin{figure*} makes the figure take up the entire width of the page
 \includegraphics[width=0.98\linewidth]{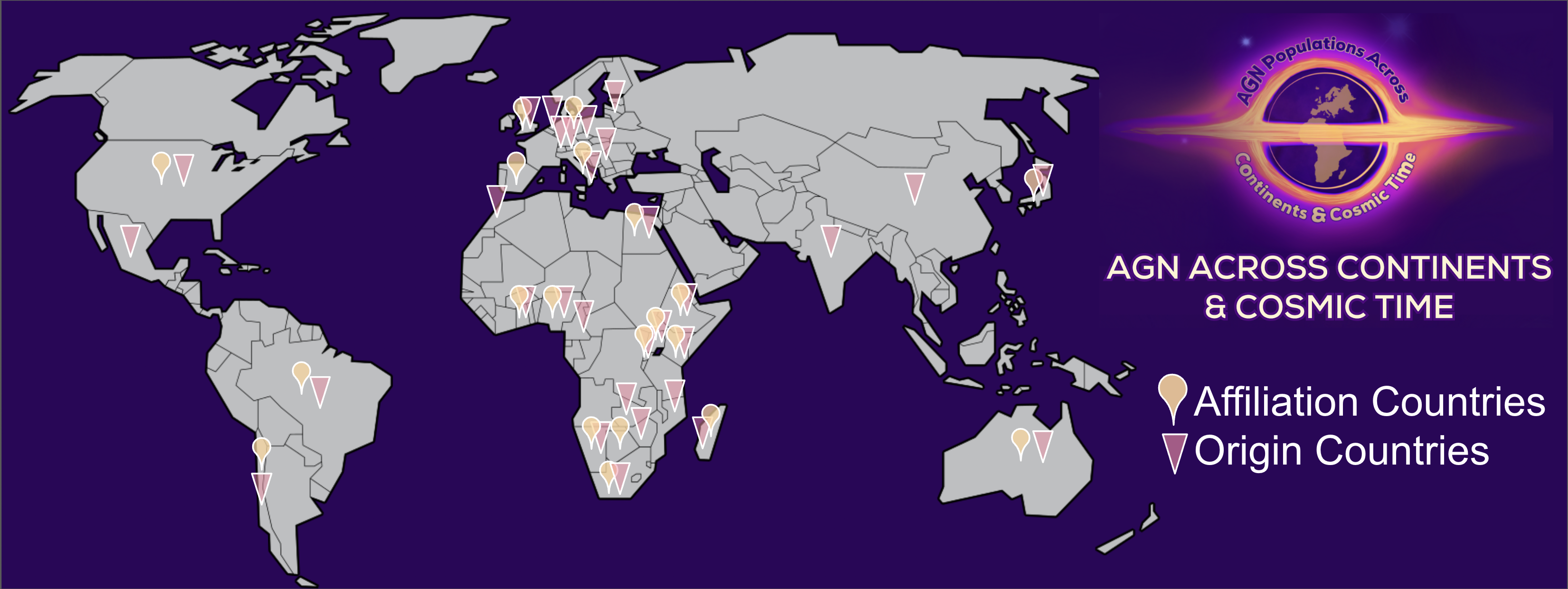}
	\caption{Map showing the countries that were represented by the participants' affiliations and origin.}
	\label{fig:particpants}
\end{figure*}

\subsection*{Selecting the LOC, SOC and Invited Speakers}
Our Local Organisation Committee (LOC) was put together primarily from the research teams (i.e., postdoctoral researchers and PhD students) of the workshop co-chairs (i.e., Chris Harrison, based at Newcastle University and Leah Morabito, based at Durham University). Therefore, they had a direct scientific interest in the workshop. The LOC were responsible for the typical activities of organising a conference (e.g., coordinating with participants, helping with administrative tasks, supporting the activities during the week etc.). Within the LOC there were four researchers with an African origin (representing four different African countries), although they were based at UK institutions at the time. Their perspective was crucial for helping to identify - and overcome - many of the challenges that African participants would face in travelling to the UK, as well as for developing a programme of activities to ensure participants felt included, supported, and comfortable to contribute. A photograph of the LOC is shown in the top panel of Figure \ref{fig:soc_loc}. 

The Scientific Organising Committee (SOC) were responsible for designing the scientific programme, choosing invited speakers and selecting the conference participants. We ensured that we had significant representation of scientists at African institutions as well as representatives from the UK and wider Europe. Of the nine SOC members, four were based at African institutions (from three separate countries). Furthermore, the selection of the SOC was made to ensure representation of established experts covering a diverse range of the scientific topics that were the focus of the conference. Consequently, the SOC not only provided crucial scientific input, but additionally were able to help design a participant selection process that was suitable for African-affiliated researchers (especially early-career researchers). Furthermore, they suggested activities for the programme that would be inclusive to those who have little-to-no experience of international conferences. Importantly, these SOC members could locally advertise the conference across the African community. A photograph of the SOC is shown in the bottom panel of Figure \ref{fig:soc_loc}. 

The SOC chose to select only a small number of invited speakers; enough to highlight the programme with some internationally-known experts in the field, but not too many as to take significant space in the programme away from contributions by regular participants. The SOC's aims of the invited speaker selection were to select four established experts who would be able to provide an introductory review talk across the four main themes of the conference. Further requirements were to include at least one speaker of African origin and affiliation, having representation of more than one gender across all the speakers, ensuring the speakers would provide a talk at an accessible level to a broad range of career stage participants, and only selecting speakers willing to actively engage with the broader development and networking goals of the conference.

\recommendation{Key Recommendation: a cross-community perspective should be considered a requirement of any LOC and SOC for any conferences with similar goals to ours, relating to bringing together research communities with diverse cultures and experiences.}

\subsection*{Selecting participants}
We planned for $\approx$110 conference participants. This number was small enough to facilitate productive discussions and effective networking, but large enough to have a number of participants from across different continents and career stages. We aimed for a fully in-person conference, to maximise the benefits of social interaction, building relationships, and ad-hoc informal discussions. Partial online participation was made available for participants who had challenges and could not travel (e.g., personal circumstances or visa complications). 

To achieve our objective of a high fraction of participation from African astronomers, it was crucial to provide full funding to those who do not have access to any travel funds (the majority of the African-affiliated astronomers). We estimated an average cost per fully-funded participant travelling from Africa, to be $\approx$£1900, to cover: travel, visa costs, accommodation, travel insurance, and local subsistence (including all meals). Our available budget limited us to fully-fund $\approx$35--40 participants, depending on exact final costs per participant. 

Throughout the selection process we kept track of both the required funding to support African participants and the participant demographics. For the demographics, we focused on a proactive approach to ensure a high level of representation of those who are African affiliated and/or have African origin. We also kept track of the distribution of career stages and gender; however, our selection approach naturally led to good representation across these two characteristics, and required no positive actions to address the diversity.

Along with a dedicated website, we released an application form for conference participation ten months before the actual conference. The deadline for submitting an application was around seven months before the conference dates. This long lead time was critical to ensure sufficient time to support travel and visa applications, as well as plan and to conduct pre-conference activities (described later). 

We aimed to include participants who were dedicated to contribute to, or benefit from, the wider networking aspects of the conference. Therefore, we did not exclusively request abstracts for talks/posters in the application form. We concluded that requiring a high-quality scientific contribution (talk/poster) abstract to attend could bias against those who had so far had limited opportunities to work on significant, internationally cutting-edge research projects. Therefore, we had three separate categories in the application form, which were used for selecting participants. These were:
\begin{enumerate}
    \item A standard scientific abstract for a contributed talk and/or poster. 
    \item Proposal for a collaborative research project, which could be discussed during the conference.
    \item A statement of interest to attend. 
\end{enumerate}
This final category was designed to capture those who may not yet have established research outputs. For example, PhD students who were at the beginning their academic studies but were strongly motivated by the opportunities of the conference. Applicants could apply using one, two, or three of these categories. These three categories were each treated separately. In effect, every applicant had up to three possibilities to be selected.

We received a total of 249 applications (once removing clearly inapplicable or duplicate applications). Of these, 75 were from people with an African origin and 74 were from African affiliations (not mutually exclusive). 

The SOC scored the submitted applications with a blinded approach (i.e., with no information on the applicants' names, affiliations etc.). Every SOC member assessed every submission for application categories 2. and 3. from the above list. However, depending on the scientific expertise of the SOC member, they only assessed abstracts for talks/posters based on the most applicable scientific theme for their own expertise. 

Each SOC member scored the submissions with an integer grade from 1 (low) to 5 (high). The scores for each submission were then collated and averaged. 

The first pass of participant selection was to take a simple cut in the average score and select everyone with a score above some fixed value. These values were chosen to select the number of contributed talks we aimed for in the final programme (i.e., 47), and then fill the remaining participant places with those who scored highly in the other two submission categories. However, this completely blind and uniform approach to selection was insufficient to obtain the desired representation of African participants. On average, the African submissions scored $\approx$0.6 points lower than other submissions. This gap is unsurprising due to the generally lower levels of available mentoring and peer support (including for tasks such as improving conference abstract drafts) compared to more deeply established research environments in places like the UK. 
We therefore decided to apply a positive action by up-weighting scores from South African submissions by 0.5 and other African submissions by 0.8. This divide within the African submissions into South Africa and other countries, was because the South African astronomy community is rapidly growing and applicants there tend to have larger peer-to-peer support groups and more mentoring, likely contributing to the typically higher submission scores. This approach ensured good geographical representation from across the continent. 

We had capacity for 120 participants and an ideal target of $\approx$110. The final number of participants attending the conference was 107, of which nine could only join online due to not being able to travel after unforeseen circumstances. Figure~\ref{fig:particpants} shows a world map highlighting the countries of origin and affiliation that were represented across the conference participants. Overall, 43\% of the participants were either from an African institution and/or of African origin. There were 20 different affiliation countries represented, of which 11 were from Africa. The break down of career stage was: 31\% senior academic/lecturer; 26\% postdoc/fellow; 31\% PhD student; and 12\% Masters students. For gender, the final breakdown was 49\% Male, 49\% Female, and 2\% non-binary. A conference photo, including the majority of participants, is shown in Figure \ref{fig:conference_photo}.

\recommendation{Key Recommendations: A long lead time for opening applications is crucial. Multiple categories for applications to attend (not just scientific abstracts) can help increase the diversity of the participants. Initial selection should be blind (anonymous) and based on a scoring rubric but positive action might be required to achieve the objective diversity.}

\subsection*{Financial and practical support of participants}
Offers for conference participation were sent out roughly six months before the conference itself. This long lead-time was chosen to ensure sufficient time to help participants who needed to apply for visas, and/or whose travel we needed to support. Indeed, visa applications can take a significant amount of time.  For example, one participant had a visa issued only two days before travel. For similar reasons, it was necessary to book travel for fully-funded participants \textit{before} visa applications were completed, otherwise flight costs could have risen to unaffordable rates as the travel dates approached. 

Of all participants, 38 received full funding from our budget, with a small additional number of participants receiving part funding based on a case-by-case assessment of need (e.g., fee waivers for early-career researchers from any institution/country without large funding resources). As many African-affiliated participants had no means to make significant upfront costs, it was necessary for the conference host university to book the travel and the accommodation directly on their behalf. Due to the nature of the accommodation we used (university colleges), it was also possible to purchase all evening meals for the fully-funded participants in advance of the conference. Fully-funded participants also had their posters printed (when relevant) directly at the conference venue, at no cost to themselves.

The only costs that fully-funded participants were required to pay for themselves up-front, and later obtain a reimbursement, were the visa costs, travel insurance (as often not provided by the home universities), and local travel costs. We note that costs associated with obtaining visas were sometimes quite high, due to participants needing to travel to consulates (sometimes multiple times). Although this process generally went well, some applicants had no access to personal bank accounts and arrangement had to be made with a third party to make the reimbursement because cash payments were not possible.  In some cases, we were not aware of this until after it was too late to help with alternative methods of payment beforehand.   

We supported visa applications by providing the necessary invitation letters to attend the conference. We provided ad-hoc help for some participants on how to complete visa applications. However, with hindsight, we should have provided more specific guidance for those who had not previously applied for visas. For example, some visas were initially rejected on the grounds of not providing sufficient evidence that they had reasons to return home after the conference (e.g., by providing letters from their home universities to prove employment or registration on PhD programmes). 

The participants applied for, and received, standard visitor visas. These are valid for six months, and some participants organised other work visits while they were in the UK. This required some extra coordination, but we were happy to support it. We had some requests during the meeting, or after, for other work visits and it would have been useful to highlight this opportunity to participants beforehand. This could have allowed participants to make the best use of their visas while they were valid. However, at the time, we caution that while visits to the UK for conferences and meeting did not require approval from the Academic Technology Approval Scheme (ATAS), research visits of any length did require approval. If people wish to join a visit to carry out research along with this, we recommend they apply for ATAS certification, as required, at the same time or even before the visa.

Although the LOC communication to participants directed them to ``contact us if you have any questions or problems'', we found that some did not do so until after they completed a task. In some cases (e.g., reimbursement to those who did not have bank accounts), this made fixing the problem more challenging. This may be attributable to the lack of experience or confidence of the participants. Therefore, we would strongly encourage participants to be very communicative at each stage of their diverse visa and travel processes, to give organisers a chance to assist as early as possible

\recommendation{Key Recommendations: To aid successful visa applications, ensure a long lead time and provide extensive supporting documents and guidelines for participants. Reimbursement factors should carefully be planned with local administrators, and {\em pre-payment} of all the funded participant's major expenses are crucial. Be very explicit with participants which costs they can, or can not, claim back in advance of any payments being made (e.g., if taxis can be used). Plan for extra costs associated with visa applications and travel insurance. Throughout, clear communication is critical between participants and organisers.}

\section*{Programme of Activities \& their Evaluation}

We conducted a variety of activities before, and during, the conference to aid the sharing and discussion of research outputs, as well as facilitate networking and collaborations.  To assess the success of the activities, we conducted a participant evaluation questionnaire. Participants were asked to complete this during the final plenary session itself, to help increase completion rates. Overall, 73/107 participants (68\%) completed the evaluation form. This was a representative sub-sample of all attendees, with $\approx$30\% from each of the groups of PhD students, Postdocs/Fellows, and Lecturers, and with $\approx$12\% from Masters students. Furthermore, 47\% completing the survey were from an African institute and/or African origin. We provide some key results of this evaluation alongside the relevant activity description. For the quantitative questions, we split the results by all participants and then just the African affiliated/origin participants, to assess if there were any deviations for this specific group for whom there was a focus on providing an inclusive, welcoming and collaborative environment. 

The following sub-sections describe the activities and their impact, as much as we are able to assess at the time of the workshop. 

\subsection*{Pre-conference communications and activities}
As is increasingly common practice for scientific workshops, we created a code of conduct. All participants were made aware that this applied to all activities before and during the workshop, and this was posted on our website. 

Around one month before the workshop, we created a collaborative online Slack space, to allow participants to start introducing themselves to each other. Additionally, to encourage participants getting to know each other before the conference, we asked participants to create two minute introductory videos. These videos were shared on a private YouTube playlist in the month leading up to the conference. Participants were provided with clear guidelines on how to record these videos on a phone, and what these videos should include (i.e., their name, affiliation, research interests, and what they hope to get out of the conference). Only 35/107 participants created these videos; with mostly the more junior participants participating (the LOC and SOC were all compelled to make a video). Nonetheless, we had several positive comments during the feedback questionnaire about the introductory videos being a good idea, and useful to start to get to know fellow participants. 

The most successful pre-conference networking activity was the ``Buddy'' system that was created to pair all African-affiliated early-career researchers with an non African-affiliated early-career researcher. Although this was an opt-in activity, nearly all early-career researchers participated. Buddys were put into contact and encouraged to email/call before the conference to get to know each other, and then to meet each other at the start of the conference. They were also encouraged to provide feedback to each other on their scientific contributions. The comments we received demonstrated this was extremely helpful for feeling prepared for the conference and to make participants feel more confident to integrate and participate in discussions, because they already knew somebody attending the conference. 

Due to the wide range of experience of the participants, where some may not have presented at, nor even attended, an international conferences before, we decided to provide extensive support and guidelines for preparing high-quality scientific contributions. This included: (1) an online workshop on preparing a good scientific talk; (2) guidelines and template examples on preparing a good scientific poster; (3) guidelines and examples for preparing a 90\, second `sparkler' talk (more details below). The scientific talk workshop had over 30 attendees (with a similar number watching the recording later on) and the feedback from participants highlighted that the guidelines, examples and templates, were appreciated and extensively used. We believe this pre-workshop guidance and training significantly contributed to the consistently high-quality scientific contributions delivered during the conference itself. 

\recommendation{Key Recommendations: Pre-conference networking activities and training/guidance are extremely valuable activities. These help ensure an effective meeting where participants immediately feel confident and welcome, and deliver high-quality scientific contributions.}

\begin{figure*}[ht]\centering % Using \begin{figure*} makes the figure take up the entire width of the page
	\includegraphics[width=0.45\linewidth]{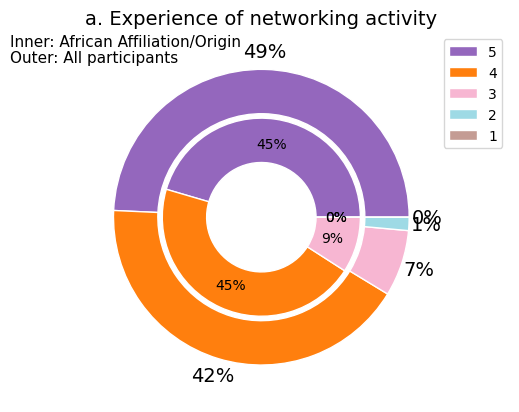}
 \includegraphics[width=0.45\linewidth]{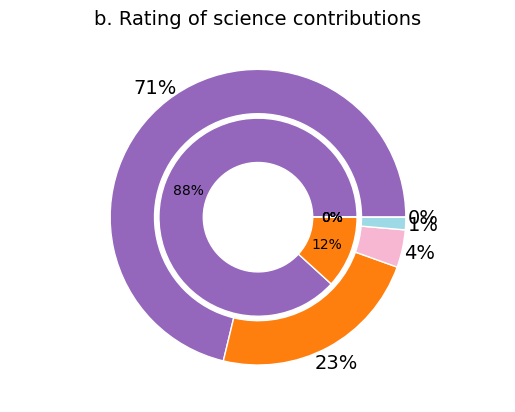}
  \includegraphics[width=0.45\linewidth]{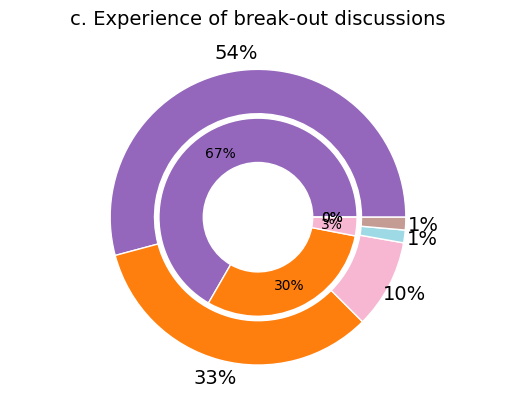}
    \includegraphics[width=0.45\linewidth]{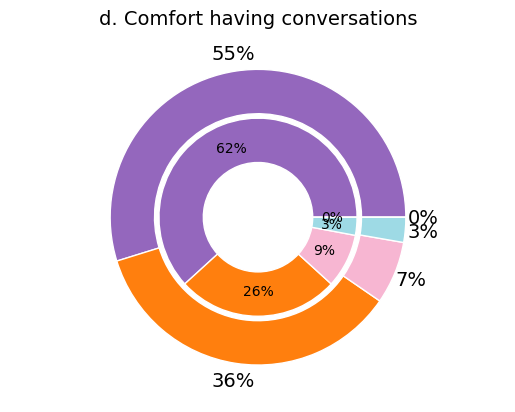}
	\caption{Results from four of the rating questions of the feedback questionnaire, where 5 is high and 1 is low. The outer circles are for all participants, and the inner circles for those with African affiliation/origin. The questions were: a. \textit{``How did you find the networking session on Monday?''}; b. \textit{``Rate your overall experience of participants' science contributions part of the conference (contributed talks, sparkler talks, posters)''}; \textit{c. ``How did you find the break-out/discussion sessions throughout the week}''; \textit{d. ``How comfortable did you feel to have discussions/conversations with new people during the conference (e.g., during coffee breaks, lunchtimes...)''}.}
	\label{fig:feedback}
\end{figure*}

\subsection*{Day 1 Networking activity}
After the first welcome and introductory talks on the first day of the conference, we conducted a 1.5\,hour networking/ice-breaker activity. We felt this important to create a welcoming atmosphere where people immediately felt comfortable to approach others for later scientific discussion. Therefore, the focus was on getting to know each other, and less on discussing scientific topics. 

We split into groups of $\approx$25, across separate rooms. These groups were then re-shuffled half way through the session. Before the activity all participants had been assigned their groups/rooms so that they knew where to be. Two activities were used: (1) ``Find 10 things in common'' and (2) ``Queen Bee: Find the Answer to my Questions''. The former activity had each group split into 2-3 sub-groups and then take turns to briefly introduce themselves to each other. They then had ten minutes to try and find ten things in common across the whole group. This could be related to research tools used, hobbies, favourite types of food etc. The second activity involved sub-groups of $\approx$3--5. One person in each group is nominated the ``Queen Bee'', who gives their ``worker bees'' a question that they want the answer to. This could be something scientific (e.g., ``who could I talk to about processing these type of astronomical data''), or anything else (e.g., ``what's the best restaurant to get good Indian food in Newcastle''). All the worker bees, from each group, then spend roughly three minutes speaking to everyone else in the room to get the answers, and report back to the Queen Bee. The Queen Bees could then report back to the whole room what they found out, before somebody else takes a turn as Queen Bee.

The top-left panel of Figure~\ref{fig:feedback} shows the results of asking participants ``How did you find the networking session on Monday?'', with answers between 1 for poor and 5 for excellent. It can be seen that the activity was extremely popular across all participants, with 91\% scoring 4 or 5. In Figure~\ref{fig:feedback} we also split the responses for only the African affiliated/origin participants, and found equal levels of positive responses for this specific group. The qualitative comments suggested the activity was very effective for helping early-career researchers feel more relaxed and able to approach people for conversations throughout the week. It was especially noticeable feeling more comfortable approaching senior academics after this activity. Some other comments suggested the activity was a bit tiring, so it could have been reduced in length and/or more rest time built in to the programme after this intense activity. 

\recommendation{Key Recommendations: A networking / ice-breaking session can be very effective to enable participants to feel relaxed and more able to approach other participants to initiate discussions throughout the conference.}

\subsection*{Scientific programme}
The scientific programme included a mixture of participant contributions and break-out discussion sessions. The programme also included lengthy coffee/tea breaks (typically 45 minutes) and lunch breaks (typically 90 minutes), which helped ease any time pressure in the schedule and allowed people to spend more time networking. Quite a few of the qualitative responses to our feedback survey indicated that people appreciated this chance to talk to other participants in a leisurely way. 

\subsubsection*{Science contributions}
There were three types of scientific contributions during the workshop: (1) long-form talks (4 invited talks with 30 minute slots and 47 contributed talks with 15 minute slots); (2) posters and (3) sparkler talks. The invited and contributed talks were selected during the conference selection process, described earlier in this report. All other participants were invited to present a poster and/or sparkler talk. The posters were displayed throughout the conference in the same space as coffee and lunch breaks, which was effective for increasing activity and discussions around the posters. 

Sparkler talks were 90 second presentations (with one slide only). These were done over two scheduled blocks on the first two days. The purpose of these were to allow participants to introduce themselves, their main research interest and/or one key scientific result, and to advertise a poster (if relevant). As described above, participants were given examples and training on how to effectively put these together, with a focus on avoiding participants trying to give a full scientific talk very quickly!  

The SOC selected chairs for the contribution sessions, to be more senior academics, but keeping the geographical representation the same as the overall conference. We provided the chairs with moderator guidelines, which most chairs adhered to. These included information about keeping to time, encouraging a diversity of participants to ask questions, and ensuring the code of conduct is upheld. The LOC provided technical support for each session, with specific LOC members identified for each session.  

To facilitate early career engagement, we handed out a conference sticker to any early-career researcher who asked a question to a speaker during any of the sessions. A prize was promised to anyone who earned more than 10 stickers, and at the end of the week there were two PhD students who had won a prize. This was a very effective way to help shift the question and answer sessions more towards early-career researchers, and we highly recommend doing this at any conference. 

The top-right panel of Figure~\ref{fig:feedback} shows the results of the participant feedback to the question: ``Rate your overall experience of participants science contributions part of the conference (contributed talks, sparkler talks, posters)'', ranging from 5 (Excellent) to 1 (Poor). The results are very positive with 71\% ranking 5 and 23\% ranking 4 across all participants. For only the African affiliated/origin, the results were 88\% ranking 5 and 12\% ranking 4. This highlights the high-quality scientific contributions and that we achieved our objective to put on a high-quality scientific meeting, addressing relevant problems of the field. 

\subsubsection*{Break-out discussions}
During the week we organised three break-out discussion sessions, of 90 minutes each. We had four rooms available for parallel discussion sessions. On the first session, we had a pre-planned 45\,minute panel discussion with all participants. However, the rest of the discussion time was not pre-planned. Instead, participants were encouraged to choose topics of interest that they wanted to discuss and self-organise discussion sessions across the available rooms. The proposals for collaboration research topics that were submitted at application were also shared among participants at the beginning of the week, to help people find researchers with similar interests. This organisation was done via a collaborative, online document, so people could assign discussion topics to different rooms and/or decide which discussion topics to join. Online participants were also encouraged to participate in a hybrid format; however, we found limited engagement from online participants in this activity. 

The bottom-left panel of Figure~\ref{fig:feedback} shows the results of the participant feedback to the question: ``How did you find the
break-out/discussion sessions throughout the week'', ranging from 5 (Excellent) to 1 (Poor). The feedback was broadly positive with 54\% of participants scoring 5 and 33\% scoring 4. The qualitative comments imply that the overall concept and freedom to self-organise was appreciated. It was also appreciated that significant time was scheduled for discussion sessions, rather than the programme being dominated by presentations. However, the comments also implied that an improvement would have been to find methods for the less confident participants to participate in the discussions themselves (which were often dominated by more senior academics, often of non-African affiliation/origin). An example might have been using online voting tools, or collaborative documents, for people to provide their input into the discussions, but in a non verbal way.  

\recommendation{Key Recommendation: Engaging early with all participants to offer training for giving talks can be key to ensuring a high-quality scientific meeting. Time for discussion is valuable, with the freedom to let participants chose their own topics and groups. It is important to find methods, including incentives, for early-career researchers to more actively participate in discussions.}

\subsection*{Social events programme}
In addition to the scientific programme, we had a social programme of: (1) a welcome drinks reception on the first evening; (2) a conference dinner; and (3) a mid-week, half-day sightseeing in the city of Newcastle. These activities proved to be very popular, and a highlight was encouraging participants to wear dress relevant to their cultural background for the conference dinner. Many participants did wear cultural dress, which sparked plenty of discussion and a mini-photoshoot!

Another point of note was that alcohol was not provided during the social events (although a bar was available during the conference dinner). This decision was made to remove the pressure of the social events evolving around drinking alcohol, which might not be comfortable for some participants - as one person said in their feedback: ``Although I like alcohol, I thought it was really nice that none of the socials centred around alcohol. That way everyone felt included but those who wanted to drink could choose to.'' There were hardly any negative comments aside from a note about the poor British weather during the sight-seeing! 

\recommendation{Key Recommendation: A good social programme allows for some decompression from the intense scientific programme. Allowing the diversity of cultures to be shared and showcased during the social events can be a popular highlight and another way to build relationships and understanding among communities.}

\subsection*{Overall participant experience and feedback} 

The feedback we received indicates that we achieved our objective of creating a collaborative environment where people ``across continents'' could come together to network and share in scientific discussion. The bottom-right panel of Figure~\ref{fig:feedback} shows the results of the participant feedback to the question: ``How comfortable did you feel to have discussions/conversations with new people during the conference (e.g., during coffee breaks, lunchtimes...)'', ranging from 5 (Very) to 1 (Not at all). The results are overall positive with 55\% scoring 5 and 36\% scoring 4, with a similar positive response for only the African affiliated/origin groups (62\% scoring 5 and 36\% scoring 4). 

The quotes we received from participants during the feedback process further highlight that we achieved our objectives to put on a high-quality scientific programme, whilst simultaneously creating a comfortable and collaborative environment. Some example quotes are:

\begin{itemize}
\item ``I loved the conference, it was a perfect balance of presentations and time to talk and get to know people.''
\item ``I am a senior researcher based neither in UK nor Africa and not yet maintaining connections with Africa. I attended this conference because of my interest in the subject, and thoroughly enjoyed the strong scientific programme and meeting African colleagues on the PhD and professor levels.''
\item ``So many opportunities for talking to people in a relaxed environment.''
\item ``This is probably the most pleasant atmosphere in a conference I have ever experienced.''
\item ``What I liked the most is interacting with people. All the conference attendees are unexpectedly nice and helpful and comfortable to talk to.''
\item ``I liked how it aimed to address the inequalities that those in African countries face and bring them to a conference like this. The organisation and communication was also outstanding.'' 
\end{itemize}

We received very little feedback from the small number of online participants. However, we know that there they mostly only participated during the scientific contributions part of the conference. Our focus was always on generating an effective in-person conference experience. Future consideration would be needed to make a more effective hybrid conference format to facilitate networking, and/or to consider an online-only equivalent.

\section*{Legacy, Future and Follow-up}

\begin{figure}[ht]\centering % Using \begin{figure*} makes the figure take up the entire width of the page
  \includegraphics[width=0.9\linewidth]{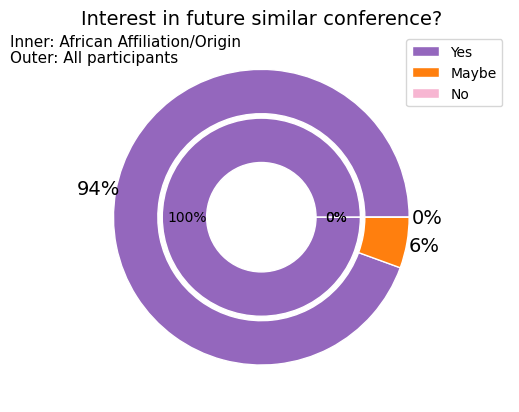}
	\caption{The results of the feedback questionnaire question: \textit{Would you be interested in attending a conference like this in a future?}, where the outer ring is for all participants, and the inner ring is for those with African affiliation or origin.}
	\label{fig:again}
\end{figure}

To provide a legacy and citable location of the scientific contributions, in particular for the early-career participants, we created a Zenodo Community\footnote{\url{https://zenodo.org/communities/agnacrosscontinents/}}. For those who wished, we added the posters and contributed talk slides to this community. 48 contributions have been added to this community. We hope that this is a useful point of reference for participants who want to cite their contribution during the conference. 

Immediately after the conference, initial signs were already encouraging for a longer-term impact of the conference. For example, when we asked ``Would you be interested in attending a conference like this in a future?'', (see Figure~\ref{fig:again}), 94\% of all participants answered ``yes'' (100\% of the African affiliated/origin participants). There was enthusiasm to run another similar meeting, potentially on the continent of Africa next time. Indeed, there are now provisional plans for an AGN themed conference to occur in Kenya in September 2026. 

%%%%%%%%%%%%%%%%%%%%%%
\subsection*{One year on: longer-term impact of the conference}

One year after the AGN Across Continents conference (i.e., in July 2025), we contacted the participants to ask about longer-term impacts that they had experienced due to the conference. The responses came in two forms: (1) non-anonymous informal email descriptions of impacts and (2) anonymous answers to a structured form (completed by 19 participants). The latter included two main parts; firstly asking the respondents to state which of the listed impacts they had experienced (see Table~\ref{tab:table1}) and secondly, open-ended requests for benefits that experienced on either research projects or their professional development. 

Our analysis finds clear evidence of long-term benefits in research collaboration, professional development, and community-building, especially for African-affiliated early-career participants. Quantitatively, a majority of respondents reported initiating new projects or ideas, while nearly all reported concrete career gains (see Table~\ref{tab:table1}). Qualitative feedback highlights multiple joint research projects launched (e.g., a new AGN program at the Ghana observatory), mentorship connections formed, and increased confidence and visibility among early-career researchers. Many participants (including one who secured a PhD supervisor from conference contacts) attribute career boosts to the conference. The ``Buddy Scheme'' introduced cross-continental mentoring, with many buddies remaining in contact, reflecting sustained networking. Feedback emphasises that greater African representation made senior researchers more approachable, benefitting students and early-career scientists alike. These outcomes align with the conference’s inclusivity goals, by \textit{``bringing researchers together from Africa, Europe, and beyond''}, the meeting fostered a more global approach to science and inclusivity.

Key metrics from the participant survey show high levels of follow-up engagement and career benefits:

\noindent\textbf{New Research Collaborations:} 58\% of respondents reported discussing project ideas with the intent to start a collaboration; 16\% have already started a specific joint project; 16\% discussed ideas but “nothing came of them” yet; and only 10\% reported no new project discussions.

\noindent\textbf{Professional Development:} All respondents (100\%) indicated that the conference gave them new research ideas, greater readiness to attend future conferences, new professional contacts, and that they listed the conference on their CV\footnote{The authors anecdotally note we have indeed seen this listed on CVs in our roles of assessing various applications.}. 74\% learned of new job or funding opportunities through the conference, and 37\% have already applied for positions they discovered through these contacts. These trends are summarized in Table~\ref{tab:table1}.

\begin{table}[!h]														
\caption[]{}

\label{tab:table1}	
\small	
\resizebox{0.5\textwidth}{!}{
\begin{tabular}{lcc}																			
\hline																			
\textbf{Research Collaboration Outcomes}	&	\textbf{Total (19)}	&	\textbf{\%}	\\
\hline	
Discussed project ideas (hoping to start) & 11   & 58\%  \\

Started a specific joint project & 3 & 16\% \\

Discussed ideas (no project started) & 3 & 16\% \\

No project discussions & 2 & 10\% \\
 \\
\hline	
\textbf{Professional Development Outcomes}	&	\textbf{Total (19)}	&	\textbf{\%}	\\
\hline	
New ideas to advance future research & 19 & 100 \\
Learned of new job/funding opportunities & 14 & 74\% \\
Applied for jobs (through contacts/opportunities) & 7 & 37\% \\
New contacts/networking established & 19 & 100\% \\ 
More prepared to attend future conferences & 16 & 84\%  \\
Conference attendance on CV & 19 & 100\% \\

\hline																	
\end{tabular}
}
\end{table}

\subsection*{Case studies of some of the outcomes}

\textbf {1. A multi-country collaboration: Pan-African AGN survey using SALT/IFS (PanAfroAGN-SI)}: Shortly after the conference, an international team including African (mostly) and European astronomers proposed a major SALT observing programme on AGN. This project built on contacts made at the workshop, and it will be led by African scientists. The programme embodies the conference theme, covering key AGN science questions, linking African telescope resources with broad expertise [7]. The project has been approved as a Large Program (3--4 years of observations), with a provisional 10 hours as a pilot conducted initially, with 100s of hours expected to follow. Data reduction of the first observations is underway.

\textbf{2. UKRI-funded GGOS-Africa project}: During the conference , the project co-ordinator (Jack Radcliffe) met future project partners for the Africa Global Geodetic Observing System (GCOS) application to build geodetic stations across Africa. The proposal was successful, funded by UKRI (with grant code UKRI476) and the project is ongoing.\footnote{\href{https://www.jb.man.ac.uk/GGOS-Africa/overview.html}{https://www.jb.man.ac.uk/GGOS-Africa/overview.html}}

\textbf{3. East African AGN Paper}: A graduate student from Rwanda combined efforts with peers in Uganda and Ethiopia to co-author a manuscript on the stellar populations of AGN. The idea emerged from networking at the conference and a follow-up virtual meeting. This new collaboration illustrates the shared expertise and data resources that participants foresaw during the post-conference discussions.

\textbf{4. DARA-STFC Postdoc fellowship placement}: A PhD student from Africa attributes their current postdoc position as a result of their ``collaborative discussions at the conference''.

\textbf{5. Strengthening African Research Capacity via ``Buddy Collaboration'':} A clear career-development success emerged from the Early-Career Researcher ``Buddy Scheme''. Through the buddy pairing, Ghanaian researcher Dr. Proven Emmanuel established a collaboration with Dr. Evaristus Iyida, resulting in a DARA postdoctoral appointment at the Ghana Radio Astronomy Observatory, significantly contributing to the institute’s scientific output and its capacity for AGN-focused research. This case highlights how intentional mentorship and cross-continental networking directly support institutional growth and long-term skills development in the African astronomy community.

\textbf{6. Research visit to Ghana by a senior European-based professor:} following discussions during the conference, a senior professor, Leah Morabito, was invited to deliver DARA lectures at the Ghana Radio Astronomy observatory. The local researchers at the observatory note that the involvement of this professor has improved the quality and the delivery of these astronomy lectures significantly.

\textbf{7. Durham visit (mentored research experience)}: Another early-career researcher arranged a three month research visit to Durham University, UK, to learn radio techniques. They attribute this to an introduction made by a conference presenter. The visit included attending a regional workshop in South Africa, strengthening "within Africa" ties. The participant reported that discussing a potential research visit at the conference put them in contact with a supportive mentor, directly benefiting their PhD research trajectory. A similar arrangement is being made for another participant to visit Durham University in early 2027.

\textbf{8. Support of masters/undergraduate transitioning to higher degrees}
The authors have a couple of examples of how junior conference participants have been supported in their transition to higher degrees, thanks to the connections made at the conference. For example, one of the African conference participants is now studying at Durham for an Master of Science by Research, supported via DARA. Another African student was successful in obtaining a PhD in Cork after support and mentorship in both searching for opportunities and preparing the application materials. The meeting led to me establishing a new collaborative project using SDSS-V data between Dr James Aird (Edinburgh University) and Dr Eli Kasai (University of Namibia), which then led to the recruitment of a Ghanian student (Mohammed Iddrisu Nlowie) through the DARA project to complete an MSc by Research in Edinburgh.

Other smaller examples include two PhD students (one of African origin) deciding to initiate a new project combining the data of one student with the methods of the other student, and a student visit to Durham  to start a radio project (combined with a workshop). It is likely other examples occurred but we do not have complete information. 

This series of examples highlights the long-term impact such a conference can have on individuals and communities. It is still too early to tell the long-term {\em scientific} impacts of this conference, as the projects/PhD programmes etc. are only just getting started. We did not take a very systematic approach to tracking long-term impacts, so there may be more examples we have missed here. Furthermore, it is clear that mentorship from new networks/relationships is a key factor in the benefits of this workshop. Therefore, on reflection we could have done more to help sustain the ``Buddy Scheme'' and potentially other mentorship relationships after the conference. 

\recommendation{Key recommendation: build in a methodology to sustain and track impacts from the workshop on a long-term time scale and consider funding possibilities for seeding ideas developed. Consider a way to sustain longer term ``Buddy Scheme'' or new mentorship relationships after the workshop.}

%: Conduct regular follow-up surveys (as planned) to document emerging projects and co-authored outputs. Providing a platform (e.g. the existing Zenodo community) to share slides and posters will help maintain momentum. Funding bodies should consider seed grants for proposals initiated at the conference.

%ii) Enhance Career Support: Formalize mentorship (expanding the Buddy Scheme) and highlight opportunities (scholarships, postdoc positions) during and after the meeting. The high interest in job/funding info (74\% learned of opportunities) suggests that integrating a career forum or networking session with funders would be beneficial.

\section*{Summary}

The main goal of this conference was to hold a scientific meeting, discussing timely scientific problems, but with a broader impact to integrate African researchers into the international community and to strengthen ties between the European and African scientific communities. According to the feedback from one participant, they reached their scientific goals for the conference, along with the bonuses of ``1. unplanned research insights and collaboration directions ...; 2. appreciating the responsibility that comes with the privilege of being West-affiliated; 3. getting to know African astronomers who are great role models for everyone.''  From a different perspective, ``Having greater representation of Africans really made people much more approachable.'' The social activities allowed everyone to bond, while many people noted that ``The quality of the participants' science contributions has been really impressive.''  The feedback has been overwhelmingly positive, and there is appetite for another conference in the community. We seem to have struck the right balance between science and social aspects about right: as one participant noted the conference was ``really well thought out in terms of inclusivity, time table, networking opportunities, making people (especially students) feel comfortable to attend and talk to more senior researchers, super positive and supportive atmosphere, extremely interesting and high-quality science.'' 

We have learned a lot from this entire process, from simple things like: what people actually have to pay for if applying for a visa and the logistical considerations we take for granted that may be different in other countries. Even the convention of putting a first name before, or after, a family name differs across countries. The diversity of the LOC and SOC were fundamental in identifying potential issues before they arose, and problem-solving when unexpected challenges came up. The biggest challenges revolved around ensuring our budget would stretch to cover all of the costs we promised, and our key recommendation here is that funding requests start early, and that as much as possible should be paid for directly by the host organisations. 

The funding for this conference pulled together from several sources. It started from the fact that the co-chairs had small amounts from their fellowships for modest science meetings, which could be combined for a single, more impactful meeting. Other funding was necessary, and it is crucial to start looking for funding sources early ($\approx$18--24 months before).

Overall, this meeting exceeded our expectations. The LOC and SOC did an incredible job to enact the vision, and to ensure that every small aspect was considered and addressed. Before and during the conference, the engagement from all participants was absolutely amazing. It was wonderful to see senior and junior researchers from Africa, Europe and beyond, interacting and discussing science. We hope that the collaborations that have started will bring our communities together for a more global approach to science and inclusivity. 

Reflecting on the impacts we have managed to document a year after the conference, we conclude that the AGN Across Continents conference has had a measurable, lasting impact. It seeded new research collaborations, advanced careers, and strengthened a diverse, inclusive community of astronomers. The overwhelmingly positive feedback and ongoing collaborations demonstrate that the event not only met its immediate objectives but also laid foundations for sustainable African–European research network. Future initiatives should build on these successes by supporting follow-up collaborations, expanding mentoring schemes, and continuing to prioritize diversity and inclusion in academic conferences.

%iii) Plan Inclusively from the Start: Begin outreach and visa/budget preparation 18–24 months before the event (as the organizers themselves noted). Maintaining diverse local organizing committees helped identify cultural needs early; replicating this model is advised.

%iv) Build Long-Term Ties: Encourage multi-year initiatives (e.g. an African-European AGN research network or biennial conference). The overwhelmingly positive response (94\% would attend again) shows community appetite. The organizers’ aspiration that these collaborations “bring our communities together for a more global approach to science and inclusivity” is on track, and continuing support will ensure lasting impact.

%------------------------------------------------

\phantomsection
\section*{Acknowledgments} % The \section*{} command stops section numbering

\addcontentsline{toc}{section}{Acknowledgments} % Adds this section to the table of contents
Funding for the conference was supplied through: two United Kingdom Research and Innovation grants\\ (MR/V022830/1 and MR/T042842/1); the Future Leaders Fellowship Development Network Fund (PF-020); the Science Technology Facilities Council; a Royal Astronomical Society (RAS) conference/Meeting travel subsistence fund; a RAS-Astro4dev mobility grant (funded via RAS and the International Astronomical Union Office of Astronomy for Development); the Centre of Extragalactic Astronomy at Durham University and the School of Mathematics, Statistics and Physics at Newcastle University. The conference would not have been possible without the incredible efforts of the Local Organising Committee: Ivan Almeida; Emmanuel Bempong-Manful; Emmy Escott; Houda Haidar; Ann Njeri; James Petley; Shufei Rowe; and Nicole Thomas. We thank Event Durham, and in particular Thomas Ludlow, for help with all of the logistics of the conference and travel arrangements. We are grateful for the crucial role of the Scientific Organising Committee in designing and planning the scientific programme, made up of: James Aird; James Chibueze; Eli Kasai; Mirjana Pović; Cristina Ramos Almeida; Zara Randriamanakoto; and Brooke Simmons.
%----------------------------------------------------------------------------------------
%	REFERENCE LIST
%----------------------------------------------------------------------------------------

\phantomsection
%\bibliographystyle{unsrt}

%----------------------------------------------------------------------------------------

\appendix
\renewcommand\thefigure{\thesection.\arabic{figure}} 
\setcounter{figure}{0}   

\section{Photographs}
\begin{figure*}[ht]\centering % Using \begin{figure*} makes the figure take up the entire width of the page
	\includegraphics[width=0.9\linewidth]{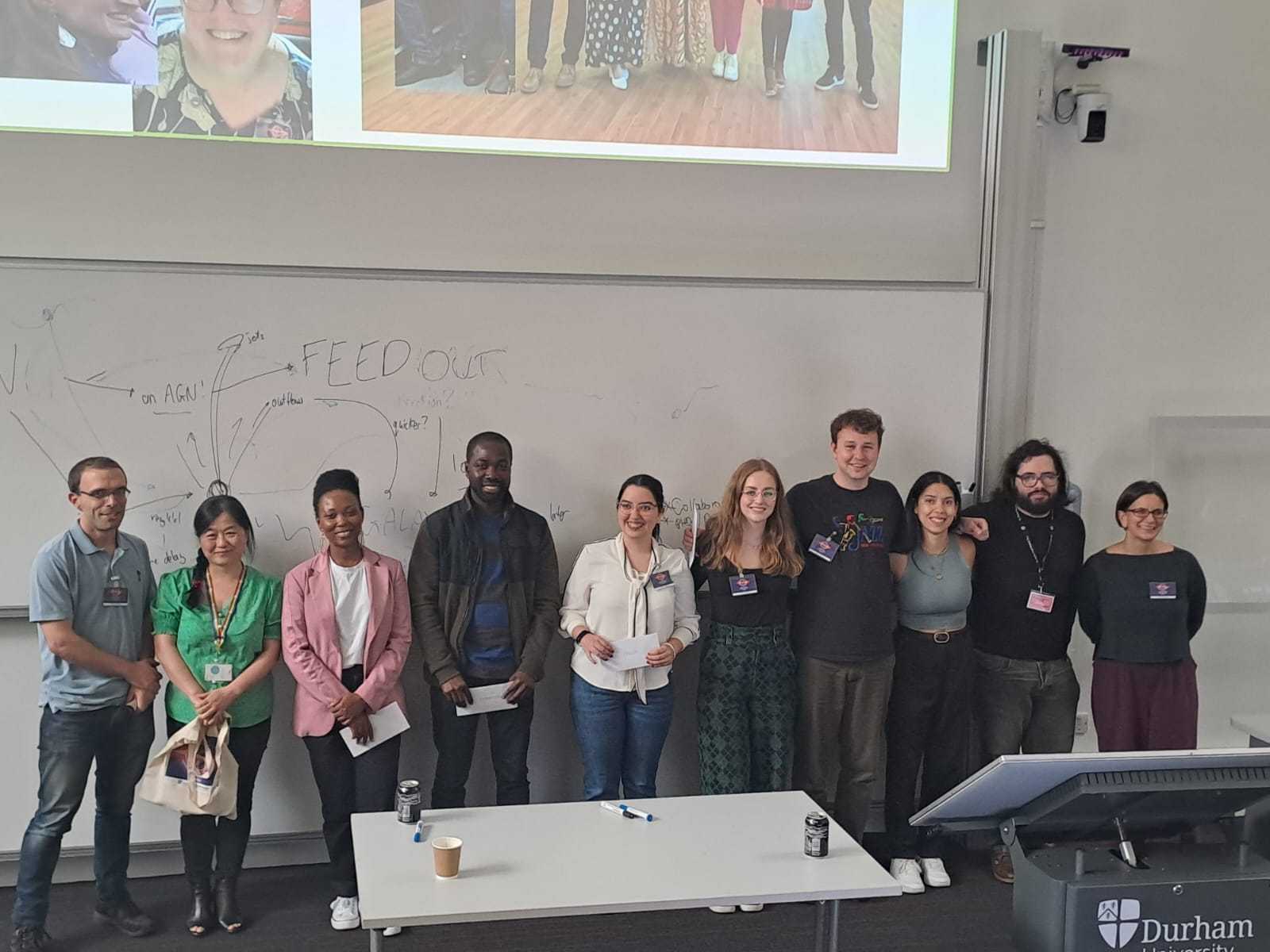}
 \includegraphics[width=0.9\linewidth]{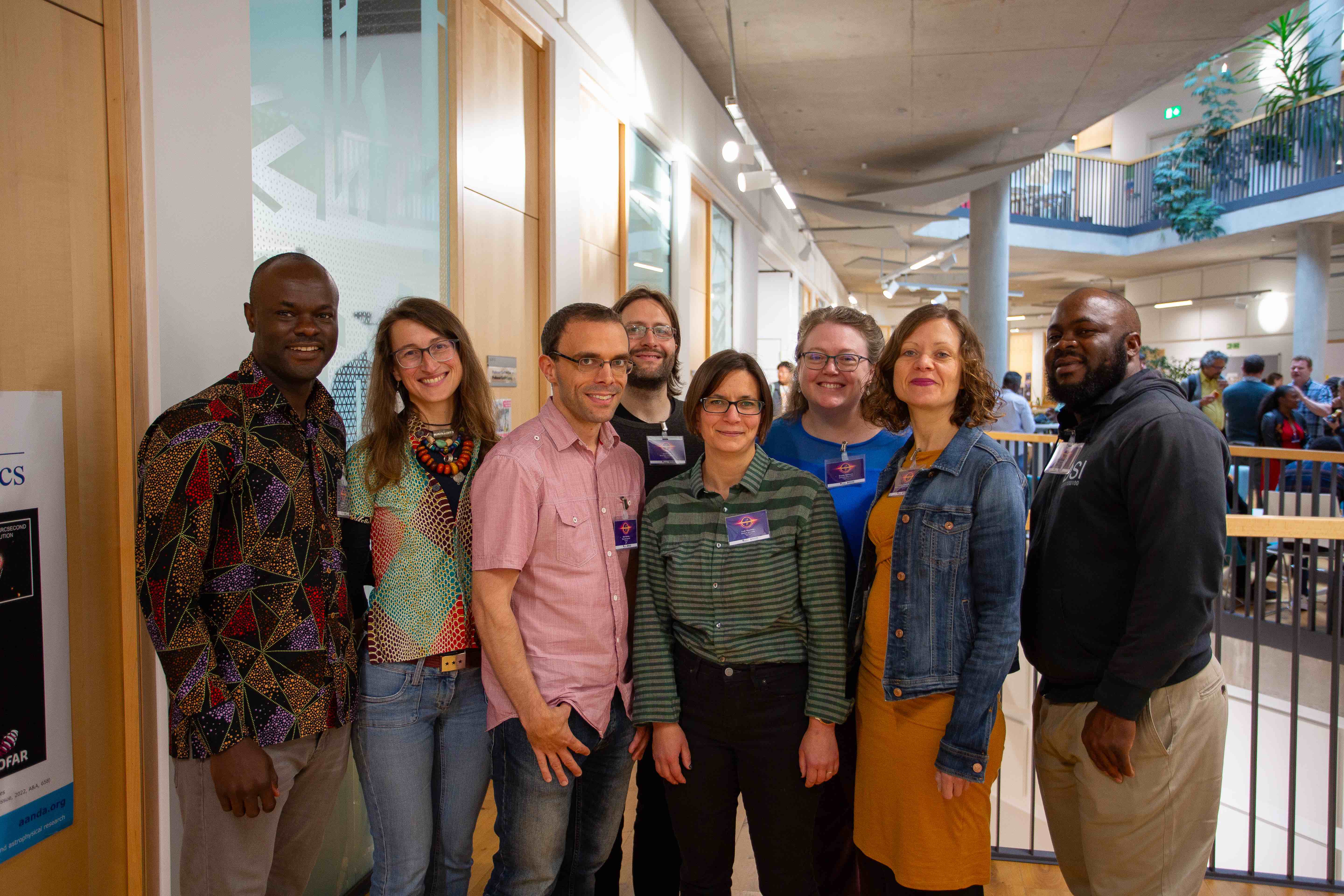}
	\caption{\textit{Top:} Local Organising Committee. From left-to-right: Chris Harrison, Shufei Rowe, Ann Njeri, Emmanuel Bempong-Manful, Houda Haidar, Emmy Escott, James Petley, Nicole Thomas, Ivan Almeida and Leah Morabito. \textit{Bottom:} Scientific Organising Committee. From left-to-right: James Chibueze, Mirjana Povi\'{c}, Chris Harrison, James Aird, Leah Morabito, Brooke Simmons, Cristina Ramos-Almeida and Eli Kasai (photo missing Zara Randriamanakoto).}
	\label{fig:soc_loc}
\end{figure*}

\begin{landscape}
\begin{figure}[ht]\centering % Using \begin{figure*} makes the figure take up the entire width of the page
	\includegraphics[width=0.9\linewidth]{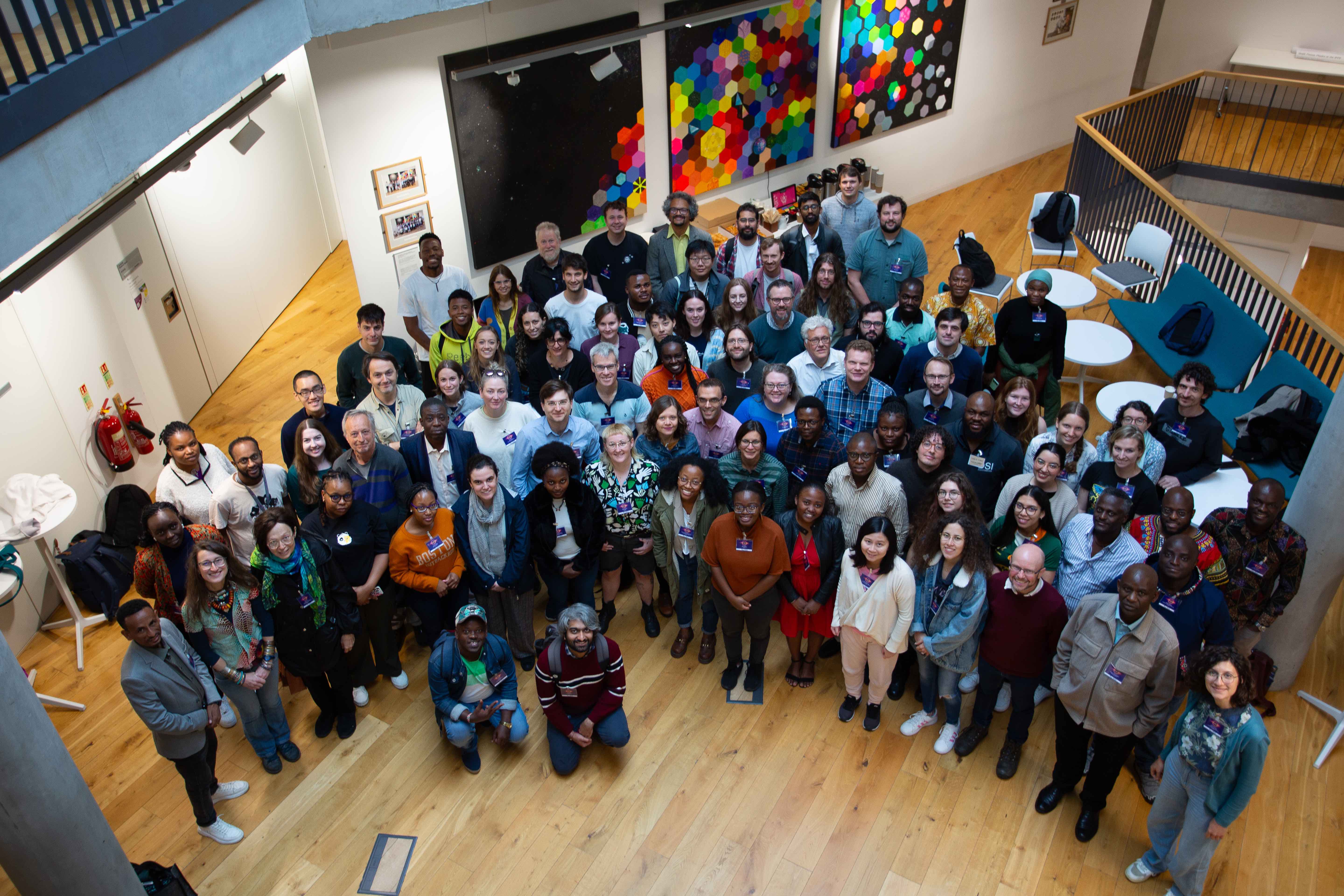}
	\caption{Conference photo, including most (but not all) participants.}
	\label{fig:conference_photo}
\end{figure}
\end{landscape}

Here we share photographs of the Local Organising Committee and Scientific Organising Committee (Figure~\ref{fig:soc_loc}) and a conference photograph of most of the participants (Figure~\ref{fig:conference_photo}).

\end{document}